\documentclass[useAMS, onecolumn, usenatbib]{mn2e}

\usepackage{xspace}
\usepackage{amsmath}
\usepackage{epsfig}
\usepackage{defs}

\pdfoutput=0

\def\ie{{\it i.e.}}
\def\eg{{\it e.g.}}
\def\ltsima{$\; \buildrel < \over \sim \;$}
\def\simlt{\lower.5ex\hbox{\ltsima}}
\def\gtsima{$\; \buildrel > \over \sim \;$}
\def\simgt{\lower.5ex\hbox{\gtsima}}

\def\ion#1#2{\text{#1\,\sc #2}}

\def\HII{{\ion{H}{ii} }}

\def\drho{\delta \rho}
\def\dP{\delta P}

\def\vecU{{\bf U}}
\def\vecno{{\bf n}_{0}}
\def\vecdn{{\bf \delta n}}
\def\veck{{\bf k}}
\def\vecn{{\bf n}}
\def\vecu{{\bf u}}
\def\vecdV{{\bf \delta V}}
\def\vecxi{{\bf \xi}}
\def\tg{\tilde{g}}
\def\tnu{\tilde{\nu}}
\def\tn{\tilde{n}}
\def\nur{\nu_{\rm rec}}
\def\mach{{\cal M}}

\title[Rayleigh-Taylor Instability at Ionization Fronts]{Rayleigh-Taylor Instability at Ionization Fronts: Perturbation Analysis}

\author[M. Ricotti]{Massimo Ricotti$^{1,2}$\thanks{E-mail: ricotti@astro.umd.edu}\\
$^1$Sorbonne Universités, Institut Lagrange de Paris (ILP), 98 bis Bouldevard Arago 75014 Paris, France\\
$^2$Department of Astronomy, University of Maryland, College Park, MD 20742, USA\\
} 

\begin{document}

\maketitle

\begin{abstract}
The linear growth rate of the Rayleigh-Taylor instability (RTI) at
ionization fronts is investigated via perturbation analysis in the
limit of incompressible fluids. In agreement with previous numerical
studies is found that absorption of ionizing radiation inside the \HII
region due to hydrogen recombinations suppresses the growth of
instabilities. In the limit of a large density contrast at the
ionization front the RTI growth rate has the simple analytical
solution $n=-\nur+(\nur^2+gk)^{1/2}$, where $\nur$ is the hydrogen
recombination rate inside the \HII region, $k$ is the perturbation's
wavenumber and $g$ is the effective acceleration in the frame of
reference of the front. Therefore, the growth of surface perturbations
with wavelengths $\lambda \gg \lambda_{cr} \equiv 2\pi g/\nur^2$ is
suppressed by a factor $\sim (\lambda_{cr}/4\lambda)^{1/2}$ with
respect to the non-radiative incompressible RTI. Implications on
stellar and black hole feedback are briefly discussed.
\end{abstract}
\begin{keywords}
\HII regions — radiative transfer - hydrodynamics — instabilities —  methods: analytical -- ISM: kinematics and
dynamics — ISM: jets and outflows -- quasars: supermassive black holes
\end{keywords}

\section{Introduction}

Radiation feedback from stars and black holes is an ubiquitous
physical process important for understanding the evolution of galaxies
and the growth of black holes (BHs). Radiation-driven galactic winds
may generate powerful gas outflows regulating the co-evolution of
galaxies and the supermassive black holes at their centers
\citep{King:03, Proga:00, DiMatteo:05, MurrayQT:05, MurrayMT:11,
  HopkinsQM:12}.  Rayleigh–Taylor instability (RTI) can occur in these
winds because a high-density shell is supported by a low-density fluid
against a gravitational field or because the shell is accelerating
outward \citep[\eg,][]{Chandrasekhar:61, StoneG:07}. Renewed interest
on the growth of the RTI has followed these studies because the
fragmentation of dense galactic supershells accelerated by thermal or
radiation pressure, can reduce significantly the wind power, making
this feedback loop less effective \citep{KrumholzT:12,
  JiangDS:13}. This problem has also prevented a definitive
identification of the driving mechanisms for winds observed in a
variety of systems \citep[\eg,][]{FaucherQ:12}.

Among various stabilization mechanisms of the RTI, the presence of a
D-type (density) ionization front (I-front) has recently been
considered \citep{Parketal:13}.  That I-fronts are stable to
longitudinal perturbations has long been known \citep{Kahn:58,
  Axford:64}. The stabilizing mechanism is the gas opacity to ionizing
radiation, that is typically due to hydrogen recombinations in the
\HII region. Let's consider a perturbation at the I-front displacing
the front further from the ionizing source. The resulting increase of
the column density of ionized gas between the radiation source and the
I-front increases the gas opacity to ionizing radiation (due to the
increased number of recombinations per unit time along the ray), thus
decreases the ionizing flux at the I-front acting as a restoring
mechanism of the perturbation. Similarly, for perturbations displacing
the I-front toward the ionizing source the increase of the ionization
flux at the I-front acts as a restoring mechanism of the
perturbation. Based on a simple dimensional analysis is therefore
expected that the growth rate of the RTI is reduced for perturbations
with wavenumber $k>k_{cr} \sim \nur^2/g$. Here, $\nur$ is the hydrogen
recombination rate inside the \HII region and $g$ is the effective
acceleration in the frame of reference of the I-front. However, a
formal perturbation analysis of the I-front is necessary to derive the
linear growth rate of the RTI including the effect of recombinations.

This paper presents a linear stability analysis of accelerating
I-fronts and I-fronts in an external gravitational field.  As far as
we know this is the first published analytical perturbation analysis
of accelerating I-fronts in which the RTI growth rate is derived
including the stabilizing effect of the gas opacity to ionizing
radiation (\ie, recombinations).  This work builds on previous
pioneering analytical studies, dating about 50 years ago, on the
stability of non-accelerating I-fronts. A brief historical prospective
is useful to summarize what is already known on the stability of
I-fronts.

\cite{Vandervoort:62} was the first to present a quantitative
perturbation analysis of a non-accelerating plane I-front subject to
appropriate jump and boundary conditions, assuming that the gas is
compressible and isothermal. For simplicity the author derived the
growth rate of the front perturbations in the limiting case when the
temperature of the neutral gas is negligible with respect to the
temperature of the ionized gas. The author considered the general case
of incident radiation on the I-front with arbitrary inclination, but
neglected hydrogen recombinations inside the \HII region. With these
assumptions weak-D and D-critical type fronts were found to be
unstable. \cite{Kahn:58} had previously argued that absorption of the
incoming radiation by neutral hydrogen atoms produced by
recombinations in the ablation outflow could have a stabilizing effect
because of the stronger absorption near the dimples of the front
surface compared to the bumps.  \cite{Axford:64} building on the work
of \cite{Vandervoort:62}, but assuming incident radiation
perpendicular to the I-front, presented a quantitative study that
showed that this stabilization mechanism is effective for
perturbations with wavelengths larger than the recombination
length. \cite{Newman:67} extended this analysis to strong D-type and
D-critical fronts.  \cite{Saaf:66} extended previous analysis relaxing
the simplifying assumption of a neutral gas much colder than the
ionized gas and found a stronger suppression of the instabilities for
the short-wavelength solutions previously found unstable.
\cite{Sysoev:97} provided more complete analysis and found numerical
solutions where the growth of marginally stable solutions in the limit
of cold neutral gas were unstable at longer-wavelengths for normally
incident radiation.  \cite{Giuliani:79} and \cite{Williams:02}
presented analytical calculations and numerical simulations for the
case of inclined incident radiation on the front showing that, in the
limit of neutral gas much colder than the ionized gas, essentially all
wavelength modes are unstable.

There is less work and is entirely based on numerical simulations,
considering the RTI at accelerating I-fronts. \cite{Mizuta:05,
  Mizuta:07} studied the hydrodynamic instability of accelerating
I-fronts using two-dimensional hydrodynamic simulations in which
recombinations are either turned off or included. The authors conclude
that the RTI can only grow when recombinations are turned off. In a
series of papers \cite{ParkR:11, ParkR:12, ParkR:13} focused on the
problem of gas accretion onto BHs from galactic scales regulated by
radiation feedback. In these simulations the ionizing radiation
emitted by the BH produces a hot ionized bubble preceded by a denser
neutral shell centered around the BH. The formation of the \HII region
temporarily stops the feeding of the BH. However, the gas density
inside the ionized bubble decreases with time and eventually the dense
shell collapses onto the BH producing a burst of accretion and
luminosity. These events repeat cyclically. The D-type I-front should
be unstable to RTI, with perturbations at the smallest scale resolved
in the simulations growing on timescales shorter than the period
between two bursts, but instead the front appears remarkably stable at
all scales. \cite{Parketal:13} have shown that the growth of the RTI
at the I-front is stabilized by recombinations in the ionized gas. The
RTI can grow only during a short phase of the cycle when the I-front
is accelerating during a BH luminosity burst. The authors speculate
that the RTI is suppressed for perturbations with wavenumber
$k>k_{cr}$, however a derivation of the growth rate of the RTI at
I-fronts is not present in the literature. The goal of this paper is
to fill the gap and provide such a derivation.

This paper is organized as follows. In \S~\ref{sec:lsa} is presented a
perturbation analysis of an accelerating I-front, and the
characteristic equation for the growth rate of longitudinal
perturbations is derived. In \S~\ref{sec:res} analytical solutions of
the characteristic equation in the relevant limits are presented,
providing the dispersion relation for unstable surface modes.  A
summary of the results and a brief discussion on astrophysical
applications is presented in \S~\ref{sec:conc}.

\section{Linear Stability Analysis}\label{sec:lsa}

Let's consider a plane-parallel I-front and work in the frame of
reference comoving with the I-front.  The formalism adopted in this
paper follows closely the one by \cite{Vandervoort:62, Axford:64} for
non-accelerating I-fronts with two important modifications: (i) terms
describing the acceleration ${\bf g}\equiv g \vecno$ in the frame of
reference of the I-front are introduced, where $\vecno$ is the unit
vector normal to the unperturbed front; (ii) an incompressible
equation of state for the gas is assumed (${\bf \nabla \cdot
  u}=0$). The second assumption is justified for sake of simplicity
given the increased complexity of the equations due to the additional
terms arising because of the front acceleration. This approximation
also allows to easily check the results against the well known growth
rate of the incompressible RTI (\ie, $n=\sqrt{gk}$).

The methodology of the analysis can be summarized as follows.  The
governing hydrodynamic equations are linearized to first order for the
neutral (quantities with subscripts ``1'') and ionized gas (quantities
with subscripts ``2'') separately. These equation can be solved to
derive the pressure perturbations as a function of the velocity and
density perturbations.  We consider perturbations on the surface of
the I-front in the form $\vecxi=\vecno \xi$, where $\xi=\xi_0
\exp(nt+ikx)$ and $\xi_0$ is the amplitude of the front deformation.
The unperturbed ($U$ velocity, $\rho$ density and $P$ pressure) and
perturbed ($u$ velocity, $\drho$ density and $\dP$ pressure) physical
quantities must obey the jump conditions at the I-front derived from
mass and momentum conservation and the perturbed quantities must
vanish at large distance from the front (\ie, must obey the boundary
conditions).  Imposing these jump/boundary conditions results in a
system of four linear equations in four unknowns: $u_1$, $u_2$,
$\drho_2$ and $\xi_0$ (it will be shown later that $\drho_1=0$ in
order to obey the boundary conditions at $z \rightarrow \infty$ and
$\dP$ can be expressed as a function of $u$ and $\drho$).  By setting
the determinant of the linear system to zero (to ensure non-trivial
solutions) the characteristic equation for the growth rate of the
perturbations, $n$, is derived. Positive real solutions of the
characteristic equations give the unstable growing modes of the system
in the linear regime.

\subsection{Governing Equations}

A moving coordinate frame is chosen so that the I-front appears
stationary, with z-axis perpendicular to the front and photons
incident from $z<0$ onto the I-front placed at $z=0$ in the
unperturbed state. The region $z>0$ contains the neutral gas and the
region $z<0$ the ionized gas.  The unperturbed velocity, density and
pressure in the neutral (quantities with subscript ``1'') and ionized
regions (subscript ``2'') are the constants $U_1$, $U_2$, $\rho_1$,
$\rho_2$, $P_1$, $P_2$, respectively. An incident radiation field $J$
directed along the z-axis is assumed. Given the symmetry of the system
we can consider longitudinal perturbations of the I-front aligned
along the x-axis without loss of generality, making the problem
two-dimensional.  Perturbations of the velocity $\vecU=\vecU+\vecu$,
density $\rho=\rho+\drho$, and pressure $P=P+\dP$ fields will be
considered on each side of the front. The following governing
equations will allow to write $\dP$ as a function of the other
perturbed quantities.

The linearized equations of continuity and motion for an inviscid gas
with an external or fictitious acceleration ${\bf g}$ in the
z-direction (\ie, orthogonal to the unperturbed I-front) pointing
toward the ionized gas, are
\begin{align}
\left(\frac{\partial}{\partial t}+U\frac{\partial}{\partial
  z}\right)\drho+\rho(\nabla \cdot \vecu)+\vecu \cdot \nabla \rho &= 0,\label{eq:euler1}\\
\rho \left(\frac{\partial}{\partial t}+U\frac{\partial}{\partial z}\right)\vecu &= -\nabla \dP - \drho g \vecno . \label{eq:euler2}
\end{align}
As usual, solutions of these equations are sought in the form
$A(x,z,t)=A_s(z)\exp{(nt+ i k x)}$, where $A$ is the perturbed
quantity, $n$ is the growth rate and $k$ is the wavenumber of the
perturbations along the x-axis. Equation~(\ref{eq:euler1}), assuming
incompressible gas (${\bf \nabla \cdot u}=0$) can be written as
\begin{align}
(n+UD)\drho=-u D \rho &= 0,\label{eq:euler1a}\\
D u+ik u_x &= 0, \label{eq:euler1b}
\end{align}
where $u$ and $u_x$ are the z- and x-components of the perturbed
velocity, respectively. The abbreviation $D \equiv \partial/\partial
z$ has been adopted. Similarly, Equation~(\ref{eq:euler2}) becomes
\begin{align}
\rho(n+UD)u &= -D \dP -\drho g,\label{eq:euler2a}\\
\rho n u_x &= -ik \dP. \label{eq:euler2b}
\end{align}
Combining Equations~(\ref{eq:euler1b}) and (\ref{eq:euler2b}) an
expression for the pressure perturbation is easily obtained:
\begin{equation}
\dP= -\frac{\rho n Du}{k^2}.\label{eq:dp}
\end{equation}
Equations~(\ref{eq:euler1a}) and (\ref{eq:euler2a}) must satisfy the
boundary conditions $\drho \rightarrow 0$ and $u \rightarrow 0$ as
$|z| \rightarrow \infty$. Since $U<0$ and unstable solutions with real
part of $n$ positive are sought, the solution for the perturbed
density is
\begin{equation}
\drho=
\begin{cases}
0 & \mbox{if }z>0,\\
\drho_2 \exp\left(-\frac{n}{U_2}z\right) \exp{(nt+kx)} & \mbox{if } z<0.
\end{cases}
\label{eq:drho}
\end{equation}
Hence, the assumption of incompressible gas leads to perturbations of
the density in the ionized component while the neutral gas density
remains unperturbed ($\drho_1=0$). Note that this is not the case when
an isothermal gas is considered instead.  Finally,
Equation~(\ref{eq:euler2a}) can be solved in the neutral gas and
ionized gas.  For $z>0$ (neutral gas) where $\drho=0$, the term
proportional to $g$ cancels out, thus the governing equation is
\begin{equation} 
D^2u - \frac{k^2U_1}{n}Du - k^2 u=0.
\end{equation}
The general solution of this equation is $u=u_1\exp{(p_1
  z)}\exp{(nt+kx)}$, where
\begin{equation}
p_1=-\frac{k^2
  U_1}{2n}\left(-1-\sqrt{1+\left(\frac{2n}{kU_1}\right)^2}\right)
=-\epsilon\frac{k}{\tn}f_1(\tn,\epsilon). 
\end{equation} 
Here, since $U_2<0$, the dimensionless growth rate $\tn \equiv -n/k
U_2$ has the same sign as $n$, hence $p_1 \le 0$ for unstable modes
(\ie, $\tn>0$). The parameter $\epsilon \equiv U_1/U_2 = \rho_2/\rho_1
\equiv \delta^{-1}$ is the inverse of the density contrast between the
neutral and ionized gas, and $f_1(\tn,\epsilon) \equiv
[1+(1+4(\tn/\epsilon)^2)^{1/2}]/2$.  In the limit $\tn \gg \epsilon$,
{\it i.e.}, for large density contrasts $\delta \gg 1$, $f_1 \rightarrow \tn/\epsilon$ (thus,
$p_1=-k$)). In the limit $\tn \ll \epsilon$, $f_1 \rightarrow 1$.
Thus, from Equation~(\ref{eq:dp}) the pressure perturbation in the
neutral gas is $\dP=\dP_1 \exp{(nt+kx)}$ with
\begin{equation}
\dP_1=-\frac{n p_1 \rho_1 u_1}{k^2}= -\rho_1 U_1 u_1 f_1(\tn,\epsilon). 
\label{eq:dp1}
\end{equation}
For $z<0$ (ionized gas) where $\drho \not= 0$, the governing equation is
\begin{equation}
D^2u - \frac{k^2U}{n}Du - k^2 u - \frac{gk^2}{n} \frac{\drho_2}{\rho_2}\exp\left(-\frac{n}{U_2}z\right)=0
\end{equation}
The general solution of this equation is $u=u(z)\exp{(nt+kx)}$ with
$u(z)=g_2 \exp{(p_2 z)}+\drho_2 g/(\rho_2 n \tn^2)\exp{(-nz/U_2)}$, where
\begin{equation}
p_2=-\frac{k^2 U_2}{2n}\left(-1+\sqrt{1+\left(\frac{2n}{k U_2}\right)^2}\right)=\frac{k}{\tn}f_2(\tn).
\end{equation}
Here, $f_2(\tn)\equiv [-1+(1+4\tn^2)^{1/2}]/2$, and $p_2 \ge 0$ for
unstable modes with $\tn>0$. In the limit $\tn \gg 1$ ({\it i.e.},
when $n \gg k|U_2|$, always valid for unstable modes when $U_2
\rightarrow 0$), $f_2 \rightarrow \tn$ ($p_2=k$). In the limit $\tn
\ll 1$, $f_2 \rightarrow \tn^2$.  The constant $g_2$ can be expressed
in terms of $u_2=u(z=0)$ as $g_2=u_2-\drho_2g/(\rho_2 n \tn^2)$ and using
Equation~(\ref{eq:dp}) the pressure perturbation in the ionized gas is
$\dP=\dP_2 \exp{(nt+kx)}$ with
\begin{equation}
\dP_2=\rho_2 U_2 \left[u_2 f_2(\tn)+ \frac{g}{n}\frac{\drho_2}{\rho_2}\left(1-\frac{f_2(\tn)}{2\tn^2}\right)\right].
\label{eq:dp2}
\end{equation}

\subsection{Jump Conditions at the Front}

Four additional equations are needed to close the system as the
normalization constants $u_1$, $u_2$, $\drho_2$ and $\xi_0$ are still
undetermined. The equations that can be used to close the system are
the jump conditions at the I-front: two energy conservation equations
(for the x and z direction) and two momentum conservation equations in
the z-direction for the top and bottom layers. The energy conservation
jump condition for the unperturbed front is,
\begin{equation}
\Delta[\rho U^2 + P]=0, 
\label{eq:jump0a}
\end{equation}
where the notation $\Delta[A] \equiv A_1 -A_2$ is adopted. The
momentum conservation equations for the unperturbed quantities are:
\begin{equation}
\rho_1 U_1=\rho_2 U_2 = -\mu J_0, 
\label{eq:jump0b}
\end{equation}
where $J_0$ is the number of ionizing photons that reach the
unperturbed I-front per unit area and time, and $\mu$ is the mean
molecular mass.  Let's consider, as mentioned above, the deformation
of the ionization front of the form $\vecxi=\vecno \xi_0
\exp(nt+ikx)$, representing the displacement of the front with respect
to the $z=0$ steady state position. The perturbation of the velocity
of the front is $\vecdV \equiv \partial \vecxi/\partial t=n \vecxi$,
and the unit vector normal to the front is $\vecn=\vecno+\vecdn$,
where $\vecdn= - \nabla \vecxi = -i \veck \vecxi$.  The flux at the
I-front is also perturbed due to the front deformation and the density
perturbations in the ionized gas: $J(\xi)=J_0+\delta
J(\xi)\exp{(nt+kx)}$.  Is important to consider the effects of
absorption in the \HII region of ionizing radiation are it is know to
stabilize the perturbations. The absorption of ionizing photons is
described by the equation
\begin{equation}
\frac{dJ}{dz}= - \nur \frac{\rho_2}{\mu},\label{eq:j0}
\end{equation}
where $\nur \equiv \rho_2 \alpha^{(2)}/\mu$ is the hydrogen
recombination rate in the ionized region, with $\alpha^{(2)} \approx
2.6 \times 10^{-13}{\rm cm}^3{\rm s}^{-1} (T/10^4~K)^{-0.8}$
\citep{Spitzer:62}. Allowing for perturbations of the front position
and density in the ionized gas, Equation~(\ref{eq:j0}) can be
integrated to give
\begin{equation}
\delta J(\xi) = - \nur \frac{\rho_2}{\mu} \left(\xi_0 + 2\int_{-\infty}^0 \frac{\delta
  \rho_2}{\rho_2} dz\right)+O(\delta \rho_2^2).\label{eq:dj} 
\end{equation}
The first term on the right hand side of Equation~(\ref{eq:dj})
describes the front stabilizing mechanism proposed by \cite{Kahn:58},
while is not clear a priori whether the second term due to
perturbations of the density in the ionized gas is stabilizing or
destabilizing (it will be found to be a stabilizing term).  The
linearized perturbed Equations~(\ref{eq:jump0a}) and (\ref{eq:jump0b})
are
\begin{align}
\Delta[\rho(\vecdn \cdot \vecU)\vecU]+\Delta[\drho(\vecno \cdot \vecU)\vecU]+\Delta[\rho \vecno \cdot (\vecu-\vecdV)\vecU] + 
\Delta[\rho(\vecno \cdot \vecU)(\vecu-\vecdV)] &= 0,\label{eq:jump1a}\\
+ \vecdn\Delta[P]+\vecno\Delta(\delta P)+\vecno g \int_{2}^{1} dz \drho &= 0,\label{eq:jump1b}\\
\rho (\vecdn \cdot \vecU)+\drho \vecno \cdot \vecU +\rho\vecno \cdot (\vecu-\vecdV) &= -\mu \delta J\exp{(nt+kx)}.\label{eq:jump1c}
\end{align}
Substituting the expressions for $\vecdV$ and $\vecdn$ two jump
conditions at the front and two continuity equations for the neutral
and ionized gas are obtained:
\begin{align}
\Delta[\drho U^2]+2\Delta[\rho U (u-n\xi)]+\Delta(\delta P) + g \int_{2}^{1} dz \drho &= 0,\label{eq:jump1bis}\\
\Delta[U \dP] + n \xi \Delta[P] &= 0,\\
\drho U +\rho (u-n\xi) &= -\mu \delta J\exp{(nt+kx)}.\label{eq:jump1bisc}
\end{align}
The integral in Equations~(\ref{eq:jump1b}) and (\ref{eq:jump1bis})
can be easily evaluated across the front discontinuity:
\begin{equation}
\int_2^1 dz \drho = -\frac{1}{n}(- U_2 \drho_2 + u_1 \rho_1 - u_2 \rho_2),
\end{equation}
while the ionization flux perturbation at the front location is
obtained from Equation~(\ref{eq:dj}) using Equation~(\ref{eq:drho})
for $\drho$:
\begin{equation}
-\mu \delta J=\rho_2 \nur \left(\xi_0 -2 \frac{U_2}{n}\frac{\drho_2}{\rho_2} \right).
\end{equation}
A few manipulations of
Equations~(\ref{eq:jump1bis})-(\ref{eq:jump1bisc}) give the following
system of four equations in four unknowns:
\begin{align}
\drho_2 U_2^2 - 2\mu \delta J \Delta[U] + \Delta(\delta P) - g \Delta[\rho] \xi_0 &= 0, \label{eq:jump2a}\\
\Delta[U \dP] - n\rho_1 U_1 \Delta[U] \xi_0 &= 0, \label{eq:jump2b}\\
\rho_1 u_1 &= -\mu \delta J + n \rho_1 \xi_0, \label{eq:jump2c}\\
\rho_2 u_2 &= -\mu \delta J + n \rho_2 \xi_0 - U_2 \drho_2, \label{eq:jump2d}\\
\end{align}
where the relationship $\Delta[P]=-\Delta [\rho U^2]=-\rho_1 U_1
\Delta [U]$ has been used. Finally, Equations~(\ref{eq:dp1}) and
(\ref{eq:dp2}) allow to express the terms in $\dP$ as:
\begin{align}
\Delta [\dP] &= -(\rho_1 u_1 U_1 f_1 + \rho_2 u_2 U_2 f_2)-\frac{g}{n} U_2 \drho_2 \left(1-\frac{f_2(\tn)}{2\tn^2}\right),\\
\Delta [U\dP] &= -(\rho_1 u_1 U_1^2 f_1 + \rho_2 u_2 U_2^2 f_2)-\frac{g}{n} U_2^2 \drho_2 \left(1-\frac{f_2(\tn)}{2\tn^2}\right).\\
\end{align}
In order to keep the equations concise, is convenient to work using
the dimensionless variables $\tn \equiv -n/(kU_2)$, $\tnu \equiv
-\nur/(kU_2)$ and $\tg \equiv g/(kU_2^2)$ and define the functions:
\begin{alignat}{2}
F_0 &\equiv f_1+f_2 & (\lim_{\epsilon \to 0} F_0 &= \tn/\epsilon), \\
F_1 &\equiv \epsilon f_1 + f_2 & (\lim_{\epsilon \to 0} F_1 &= 2\tn),\\
F_2 &\equiv \epsilon^2 f_1 + f_2 & (\lim_{\epsilon \to 0} F_2 &= \tn),\\
G_0 &\equiv \tg\left(1-\frac{f_2}{2\tn^2}\right) \quad & (\lim_{\epsilon \to 0} G_0 &= \tg).
\end{alignat}
For convenience the limits of these functions for high density
contrast $\delta \gg 1$ (\ie, $\epsilon \rightarrow 0$) are shown as
well.  With these definitions and substituting $u_1$ and $u_2$ from
Equations~(\ref{eq:jump2c})-(\ref{eq:jump2d}) into
Equations~(\ref{eq:jump2a})-(\ref{eq:jump2b}) gives the following
system of two equations in two unknowns ($\drho_2$ and $\xi_0$):
\begin{align}
\left\{1+f_2+\frac{G_0}{\tn}+2\frac{\tnu}{\tn}[F_1+2(1-\epsilon)]\right\}\drho_2 + k\rho_2\left\{\tn F_0 + \tnu[F_1+2(1-\epsilon)]-\tg\frac{\Delta\rho}{\rho_2}\right\}\xi_0 &= 0\\
\left\{f_2+\frac{G_0}{\tn}+2\frac{\tnu}{\tn}F_2\right\}\drho_2 + k\rho_2 \left\{\tn(F_1 -1+\epsilon)+\tnu F_2\right\} \xi_0 &= 0.
\end{align}
Non-trivial solutions for the dimensionless growth rate $\tn$ are found by setting the discriminant of the linear system to zero: 
\begin{equation}
\begin{split}
\left[(1+f_2)(F_1-1+\epsilon)-f_2 F_0\right]\tn^2\\
+\left\{[(F_1+2-2\epsilon)(2F_1-2+2\epsilon-f_2)+F_2(1+f_2-2 F_0)]\tnu - \left[G_0(F_0+F_1-1+\epsilon)-f_2\tg\left(\frac{1}{\epsilon}-1\right)\right]\right\}\tn \\
 - \left[2F_2 \tg \left(\frac{1}{\epsilon}-1\right)+G_0(F_1+2-2\epsilon-F_2)\right]\tnu+G_0\tg\left(\frac{1}{\epsilon}-1\right)=0,
\label{eq:car}
\end{split}
\end{equation}

\section{Results}\label{sec:res}

The general solution of the characteristic Equation~(\ref{eq:car}) can
be found numerically as a function of $g$, $\nur$ and
$\epsilon$. However, in the limit of high density contrast across the
I-front (\ie, $\delta \gg 1$), Equation~(\ref{eq:car}) simplifies
significantly and has an analytical solution.  The special cases of
neglecting the effective gravity (\ie, stability of non-accelerating
I-fronts) and neglecting recombinations will be considered separately
to check whether known results on the classical RTI growth rate are
recovered.  Only density contrasts $\delta>2$ (\ie, $\epsilon < 0.5$)
should be considered as this condition is necessary for a D-type front
to exist.

\subsection{Limit 1: Neglecting Gravity}

Setting both $g=0$ and $\nur=0$, Equation~(\ref{eq:car}) becomes
\begin{equation}
(1+f_2)(F_1-1+\epsilon)-f_2 F_0 =0.  
\end{equation}
The real roots of this equation are negative for any value $\epsilon <
0.4$, meaning that the I-front is stable for density contrast $\delta
>2.5$. Including recombinations (\ie, $\nur >0$) has minor effects on
the stability. This is contrary to results of \cite{Vandervoort:62}
who found unstable solutions for weak D-type fronts when neglecting
recombinations. However, \cite{Vandervoort:62} assumed isentropic
perturbations while the present result is derived in the limit of
incompressible perturbations. Strong D-type fronts, that have
$|U_2|>c_{s2}$ where $c_{s2}$ is the sound speed in the ionized gas,
are instead stable even neglecting recombinations
\citep{Newman:67}. Is likely that the unstable modes found neglecting
recombinations appear because of the effects of gas compressibility.

\subsection{Limit 2: Neglecting Recombinations}

In this subsection is shown that neglecting recombinations but
including non-zero effective gravity the RTI for incompressible gas is
recovered in the limit of large density contrast ($\epsilon
\rightarrow 0$). Setting both $\nur=0$, Equation~(\ref{eq:car})
becomes
\begin{equation}
[(1+f_2)(F_1-1+\epsilon)-f_2
  F_0]\tn^2-\left[G_0(F_0+F_1-1+\epsilon)-f_2\tg\left(\frac{1}{\epsilon}-1\right)\right]\tn
+G_0\tg\left(\frac{1}{\epsilon}-1\right)=0.
\label{eq:car1}
\end{equation}
This equation can be solved for $\tn$ only numerically, however in the
limit of large density contrast $\delta \gg 1$ ($\epsilon \rightarrow
0$) and $\tn \gg 1$, the expressions for the
functions $f_1$ and $f_2$ become: $f_1 \rightarrow \tn/\epsilon$ and
$f_2 \rightarrow \tn$.  Hence: $F_0 \rightarrow \tn/\epsilon$, $F_1
\rightarrow 2\tn$, $F_2 \rightarrow \tn$. Substituting these limits in
Equation~(\ref{eq:car}) the following 4th order equation is obtained:
\begin{equation}
\tn^4=\tilde{g}^2
\end{equation}
This equation has two real solutions $\tn=\pm \sqrt{\tg}$. Hence the
growing mode in in physical units is $n=\sqrt{gk}$ that indeed is the
well known dispersion relation for the incompressible RTI in the limit
$\delta \gg 1$. A numerical inspection of the Equation~(\ref{eq:car1})
shows that for values of $\epsilon\not=0$ a better solution of the
equation is $n=\sqrt{gk A}$, where $A \equiv
(\rho_1-\rho_2)/(\rho_1+\rho_2)=(1-\epsilon)/(1+\epsilon)$ is the
Atwood number.
 
\subsection{General Solution}

The general solution of Equation~(\ref{eq:car}) can only be found
numerically.  However, in the limit $\epsilon \rightarrow 0$ ($\delta
\gg 1$) and $\tn \gg 1$ the following simplified 4th order equation is obtained:
\begin{equation}
\tn^4 + 2 \tnu \tn^3 + 2\tnu \tg \tn - \tg^2=0.
\end{equation}
This equation has two complex solutions ($\pm \sqrt{-\tg}$) and two
real solutions:
\begin{equation}
\tn=-\tnu \pm \sqrt{\tnu^2+\tg}
\end{equation}
Hence, in physical units the growing mode has a rate
\begin{equation}
n=-\nur + \sqrt{\nur^2+g k}. 
\end{equation}
Therefore, at large scales $k \ll k_{cr} \equiv \nur^2/g$ the growth
of the RTI is $n \approx \sqrt{gk}(gk/4\nur^2)^{1/2}$, suppressed by a
factor $(k/4k_{cr})^{1/2}$ with respect to the classical RTI.  A
numerical inspection of the Equation~(\ref{eq:car}) shows that for
values of $\epsilon\not=0$ a better approximation of the solution is
$n=-\nur+(\nur^2+ gk A)^{1/2}$, where $A$ is the Atwood number.

\subsection{Discussion on the Effects of Gas Compressibility}

Let's now discuss how the assumption of gas incompressibility adopted
in this work may affect the results. The discussion will be guided by
previous results on the stability of Rayleigh-Taylor modes and
non-accelerating I-fronts assuming isothermal or isentropic
perturbations.

The effect of compressibility on the linear growth of the RTI was
considered by several authors assuming either isothermal or isentropic
equilibrium states and perturbations \citep[see ][for a
  review]{Gauthier:10}.  Compressibility modifies the RTI growth rate
with respect to the incompressible case as follows: the growth rate
decreases as the ``stratification'' parameter, $g/kc_{s2}^2$,
increases and the adiabatic indices decrease.  Stratification and
compressibility effects are more important at small wavenumbers and
the growth rates are larger when the light fluid is more compressible
than the heavy one. Compressibility effects are larger at small Atwood
numbers (\ie, low density contrast).  For the isothermal case,
compressibility stabilizes the RTI \citep[\eg,][]{Blake:72,
  Mathews:77, Shivamoggi:83, Ribeyre:04}.  However, the departure from
incompressible type behavior is small in most circumstances.  The
correction term, $g/kc_{s2}^2$, is much smaller than unity unless the
effective acceleration at the I-front is $g_{eff} \gg GM/R_s^2$. It is
easy to show that for $g=GM/R_s$ the stabilizing term is of the order
$kR_{b,in}/R_s^2$, that is much smaller than unity at wavelengths of
interest $\lambda < 2\pi R_s$, where $R_{b,in}\ll R_s$ is the Bondi
radius inside the \HII region.

The stability of isentropic non-accelerating I-fronts depends on the
recombination rate but also on the Mach number of the gas downstream
the front.  Supersonic flows (strong D-type and weak R-type) are
stable even neglecting recombinations \citep{Newman:67}. D-critical
and weak D-type fronts with Mach number in the ionized gas $\mach_2 <<
1$ ($|U_2| \ll c_{s2}$) are also stable \citep{Axford:64}.  The
stability of the front found in this paper may be understood in this
context taking the limit $c_{s2} \rightarrow \infty$, that gives
$\mach_2 \ll 1$. Compressibility effects destabilize the front for
weak D-type fronts with $\mach_2 \simlt 1$, but only on scales smaller
than one tenth of the recombination scale in the ionized gas.

In light of the discussion above, the effects of compressibility on
the stability of accelerating I-fronts should be small in many
circumstances. Compressibility effects should be most relevant for
weak D-type fronts with $\mach_2 \sim 0.5$. In this regime $\tg \sim
4g/kc_{s2} \sim 1$, and $\tn \simlt 1$. However, is unclear whether the
instability growth rate is increased or reduced because
Rayleigh-Taylor modes are typically stabilized while the I-front is
slightly destabilized by compressibility effects.

\section{Summary and Conclusions}\label{sec:conc}

The growth of the RTI in D-type ionization fronts has been
investigated via perturbation analysis assuming incompressible gas.
In the limit of a large density contrast at the ionization front the
RTI growth rate has the simple analytical solution
$n=-\nur+(\nur^2+gk)^{1/2}$. Therefore, recombinations in the ionized
gas stabilize the RTI on scales larger than
\begin{equation}
\lambda_{cr}=2\pi \frac{g}{\nur^2},
\label{eq:crit}
\end{equation}
suppressing their growth for $\lambda \gg \lambda_{cr}$ by a factor
$\sim (\lambda_{cr}/4\lambda)^{1/2}$ with respect to the non-radiative
case. Contrary to previous analysis of isothermal non-accelerating
I-fronts, the front is stable when gravity and recombinations are set
to zero, suggesting that the assumption of gas incompressibility has a
stabilizing effect.  The stabilization is very effective because in
most problems $\lambda_{cr}$ is much smaller than the scales of
interest.  For instance, for non-accelerating fronts in a
gravitational field produced by a point mass $M$ with $g=GM/r_s^2$,
where $r_s$ is the location of the I-front,
\begin{equation}
\theta_{cr} \equiv \frac{\lambda_{cr}}{2\pi r_s}=\frac{GM}{\nur^2 r_s^3} \sim 4 \times 10^{-10} \frac{X}{l},
\label{eq:thcrit}
\end{equation}
where $r_s=(3S_0/4\pi n^2 \alpha^{(2)})^{1/3}$ is the Str\"omgren
radius and $S_0=5.7 \times 10^{48} s^{-1} (M/M_{\odot}) l/X$ is the
number of hydrogen ionizing photons per unit time, expressed as a
function of the Eddington ratio $l \equiv L/L_{Ed}$ and the mean
ionizing photon energy $X$ in Rydbergs. In this example, at scales
$\theta \simgt 10^{-3}$, the front becomes unstable to RTI only for
very small Eddington ratios: $l \simlt 4 \times 10^{-7} X$
\citep[see,][]{Parketal:13}.  However, during a burst of luminosity
when the front acceleration timescale $t_{acc}=(dv/dt/r)^{-1/2}$ is
comparable or shorter than the recombination timescale, the RTI can
develop even for Eddington ratios of the order of unity. Another case
in which the front is accelerating and the RTI may develop is when the
front propagates toward regions of lower density, for instance if the
ionizing source is at the center of a halo with a gas density profile
$\rho \propto r^{-2}$ \citep{WhalenN:08}.

This analytical work has been inspired by the results of numerical
simulations \citep{ParkR:11, ParkR:12, ParkR:13} showing that
recombinations in \HII regions produced by BHs accreting from a
neutral medium have a stabilizing effect on the growth of RTI at the
I-front.  In these simulations the I-front is unstable only during two
phases of the duty cycle: (i) during bursts of accretion onto the BH,
when the outward acceleration of the front increases the effective
gravity and (ii) just before a burst when the accretion luminosity
reaches a minimum value triggering the collapse of the shell and the
\HII region onto the BH.  During the burst the front fragments on the
smallest resolved scales into knots that are optically thick, casting
a shadow before dissolving (self-gravity is not included in the
simulations). The phase when the I-front collapses onto the BH is
characterized by denser fingers of gas protruding toward the BH.  The
results of the present study are in agreement with the observed
phenomenology as discussed in detail in \cite{Parketal:13}.
Simulations of moving intermediate mass BHs with radiation feedback
\citep{ParkR:13} also show the formation of a dense shell upstream the
BH supported against gravity by less dense hot gas in a
cometary-shaped \HII region. Also in this case RTI at the front
appears to be stabilized by recombinations.

The results of this work may also be relevant for the stability of
supershells in AGN or galactic winds.  In general, outflows or winds
can be produced by pressure gradients as a result of thermal energy
injection or transfer of momentum from the radiation field to the gas.
Momentum driven winds can be produced by Compton scattering of photons
on electrons, by photons scattered or absorbed by dust grains or
photo-ionization of hydrogen and heavy ions by UV and X-rays.  For
instance, radiation pressure on dust has been suggested as an
important feedback mechanism in regulating star formation on galactic
scales.  However, for optically thick media and Eddington ratios close
to unity (typical of many galaxies and star clusters), the gas and the
radiation field are coupled producing phenomena such as photon bubbles
and radiation RTI \citep{Turneretal:05, JaquetK:11, JiangDS:13}.  In
this regime \cite{KrumholzT:12, JiangDS:13} showed that the transfer
of IR radiation through the neutral gas triggers radiative RTI,
driving turbulence but not a wind. Although these models do not
include ionized bubbles, a realistic description of the wind involves
supershells produced by OB associations near the galactic disk that
may provide some stabilization during the initial phases of the wind
launching process.  A more direct application of the results of this
work regards a possible driving mechanism of AGN winds based on
photo-ionization of metal ions within 0.1 pc from the supermassive BH
\citep{Proga:00, DebuhrQM:12, Novak:12}.  Ionization fronts of metal
ions of different elements are located at various distances near the
AGN, and the ions may (or may not) be coupled to the hydrogen gas
trough Coulomb collisions \citep{BaskinL:12}. Momentum is deposited in
shells at each ionization fronts, producing an accelerating wind
subject to RTI. Ion recombinations should stabilize these shells at
least on scales larger than a critical value that can be estimated
using a modification of Equation~(\ref{eq:crit}) in which $\nur$ is
calculated for the appropriate metal ion.

\section*{Acknowledgments}
MR's research is supported by NASA grant NNX10AH10G and NSF
CMMI1125285. This work made in the ILP LABEX (under reference
ANR-10-LABX-63) was supported by French state funds managed by the ANR
within the Investissements d'Avenir programme under reference
ANR-11-IDEX-0004-02. Many thanks to KH Park, J. Drake and C. Reynolds
for inspiring the paper. Many thanks to S. Falle for refereeing the paper.

\bibliographystyle{/Users/ricotti/Latex/TeX/mn2e}
\bibliography{./RT-inst}

\label{lastpage}
\end{document}